\documentclass[12pt]{iopart}

\usepackage{graphicx}
\usepackage{dcolumn}
\usepackage{bm}
\usepackage[usenames]{color}
\usepackage{colortbl}
\usepackage[normalem]{ulem}
\usepackage{epsfig} 
\usepackage{epstopdf}
\usepackage{subfig}

\newcommand{\PreserveBackslash}[1]{\let\temp=\\#1\let\\=\temp}
\let\PBS=\PreserveBackslash

\begin{document}

\title[Wide bandgap materials for quantized {anomalous Hall and magnetoelectric} effects]{Highly-ordered wide bandgap materials for quantized {anomalous Hall and magnetoelectric} effects}

\author{M.\,M.~Otrokov$^{1,2,3,4}$, T.\,V.~Menshchikova$^{2}$, M.\,G.~Vergniory$^{1,5}$, I.\,P.~Rusinov$^2$, A.\,Yu.~Vyazovskaya$^2$, Yu.\,M.~Koroteev$^{6,2}$, G.~Bihlmayer$^{7}$, A.~Ernst$^{8,9}$, P.\,M.~Echenique$^{1,4}$,  
A.~Arnau$^{1,4}$, E.\,V.~Chulkov$^{1,3,4}$}
\address{$^1$Donostia International Physics Center (DIPC), 20018 San Sebasti\'an/Donostia, Spain}
\address{$^2$Tomsk State University, 634050 Tomsk, Russia}
\address{$^3$Saint Petersburg State University, 198504 Saint Petersburg, Russia}
\address{$^4$Departamento de F\'{\i}sica de Materiales UPV/EHU, Centro de F\'{\i}sica de Materiales CFM - MPC and Centro Mixto CSIC-UPV/EHU, 20080 San Sebasti\'an/Donostia, Spain}
\address{$^5$Department of Applied Physics II, Faculty of Science and Technology,
University of the Basque Country UPV/EHU, Apdo. 644, 48080 Bilbao, Spain}
\address{$^6$Institute of Strength Physics and Materials Science, Russian Academy of Sciences, 634021 Tomsk, Russia}
\address{$^7$Institut f\"ur Festk\"orperforschung and Institute for Advanced Simulation, Forschungszentrum J\"ulich and JARA, D-52425 J\"ulich, Germany}
\address{$^8$Max-Planck-Institut f\"ur Mikrostrukturphysik, Weinberg 2, D-06120 Halle, Germany}
\address{$^9$Institut f\"ur Theoretische Physik, Johannes Kepler Universit\"at, A 4040 Linz, Austria}

\ead{mikhail.otrokov@gmail.com}

\vspace{10pt}
\begin{indented}
\item[]\today 
\end{indented}

\begin{abstract}
An interplay of spin-orbit coupling and intrinsic magnetism is known
  to give rise to the quantum anomalous Hall and topological
  magnetoelectric effects under certain conditions. Their realization
  could open access to low power consumption electronics as well as
  many fundamental phenomena like image magnetic monopoles, Majorana
  fermions and others. Unfortunately, being realized very recently,
  these effects are only accessible at extremely low temperatures and
  the lack of appropriate materials that would enable the temperature
  increase is a most severe challenge. Here, we propose a novel
  material platform with unique combination of properties making it
  perfectly suitable for the realization of both effects at elevated
  temperatures. The key element of the computational material design
  is a{n} extension of a topological insulator (TI) surface by a
  thin film of ferromagnetic insulator{, which} is both structurally and
  compositionally compatible with the TI. Following this proposal we
  {suggest} a variety of specific systems and discuss their numerous
  advantages, in particular wide band gaps with the Fermi level
  located in the gap.
\end{abstract}

%
 \vspace{2pc}

\section{Introduction}

Chasing efficient ways to introduce
time-reversal symmetry breaking perturbations in a topological
insulator (TI) \cite{Hasan2010, Qi.rmp2011} without using an external
magnetic field represents nowadays a real challenge of research
activity. The strong efforts in this direction are motivated by the
possibility {to realize}  the quantum anomalous
Hall effect (QAHE) \cite{Qi.prb2006, Qi.prb2008}, a quantized version
of the AHE \cite{Onoda.prl2003}, and the topological magnetoelectric
effect (TME) \cite{Qi.prb2008, Essin.prl2009, Tse.prl2010}, where an
electric field induces a topological contribution to the
magnetization with a universal coefficient of proportionality
quantized in units of $e^2$/2$h$. A key ingredient of the QAH state
and TME, i.e.\ a time-reversal symmetry breaking, is achieved by
virtue of a ferromagnetic (FM) order (in the most simple case
--  with an out-of-plane magnetization; although the QAH state
can also be realized with an in-plane magnetization upon certain
conditions \cite{Zhang.prb2011, Liu.prl2013, Ren.prb2016, 
Sheng.arxiv2016}). FM order induces an exchange gap in the surface state of
three-dimensional TI or an exchange splitting of the 
TI thin film gap edges (or those of a trivial insulator thin film). Then, owing
to a spin-orbit coupling (SOC), an inversion of the band gap, formed
by non-degenerate, spin polarized bands can arise. Further, the Fermi
level should be tuned into the two-dimensional band gap
removing by this any surface or bulk
state contribution. By meeting all these conditions in a
two-dimensional TI (or trivial insulator) thin film, one could realize
the QAH state, in which, unlike a quantum spin Hall state
\cite{Kane-z2.prl2005, Shuichi.prl2006, Bernevig.sci2006}, only one
pair of bands is inverted \cite{Liu.prl2008}. Characterized by a
dissipationless edge mode carrying electrons of only one spin
direction \cite{Qi.prb2006, Qi.prb2008}, this unique effect represents
a promising platform for creation of next-generation electronic
devices as well as for incarnation of novel phenomena like Majorana
fermions \cite{Qi.rmp2011}.  On the other hand, if the TI film is
sufficiently thick to eliminate the finite-size effect \cite{Qi.sci2009,
  Tse.prl2010, Nomura.prl2011, Wang.prb2015}, the exactly quantized
TME is expected to appear \cite{Qi.prb2008}, its direct consequences
being the image magnetic monopole and topological Kerr or Faraday
rotation \cite{Qi.rmp2011}. While the QAH state has already been
observed \cite{Chang.sci2013, Checkelsky.natp2014, Kou.prl2014,
  Chang.natm2015, Kou.ncomms2015, Feng.prl2015, Mogi.apl2015,
  Lachman.sciadv2015, Grauer.prb2015, Chang.prl2016} and lately found
to show indications of TME in the
Cr$_{x}$(Bi$_{1-y}$Sb$_{y}$)$_{2-x}$Te$_3$ thin films
\cite{Okada.ncomms2016}, the practical realization of the TME at a 3D
TI surface without an external magnetic field still remains elusive,
although highly desirable \cite{Qi.sci2009, Tse.prl2010,
  Nomura.prl2011, Wang.prb2015}.  So far the quantized magneto-optical
effect has been observed only at the surface of a nonmagnetic TI under an
external magnetic field \cite{Wu.sci2016, Dziom.arxiv2016}.

Nowadays, that is almost four years after its realization, the
rising of the QAHE temperature above 2 K still represents a great
challenge. The main obstacles seem to be the inhomogeneity of the
topological state coupling to the doped-in magnetic moments and the
presence of the parasitic conduction in the bulk-like region
\cite{Mogi.apl2015, Chang.prl2016}. The former is clearly illustrated
by a recent scanning tunneling spectroscopy study of the single
crystal Cr$_{x}$(Bi$_{1-y}$Sb$_{y}$)$_{2-x}$Te$_3$ surface, revealing
that a random distribution of the magnetic dopants results in a
fluctuation of the Dirac point (DP) gap size, varying roughly from 10
to 50 meV \cite{Lee.pnas2015}. This diminishes the effective energy
barrier for the carriers thermal activation that determines the
temperature scale of the QAHE observation. Another possible factor
contributing to the QAH state deterioration is an appearance of the
dissipative conduction channels \cite{Mogi.apl2015} due to
metallization of the magnetically-doped TI bulk-like region.

Here, using first-principles calculations, we propose a new, simple
and efficient method to incorporate magnetism in TI surfaces that permits
to avoid problems caused by 
an inhomogeneous dopants distribution and/or possible metallization of
the bulk-like region. Instead of a surface \cite{Liu.prl2009,
  Schlenk.prl2013, Otrokov.prb2015} or bulk doping \cite{Qi.prb2008,
  Vergniory.prb2014, Lee.pnas2015}, surface alloying
\cite{Henk.prl2012.2, Schlenk.prl2013, Polyakov.prb2015} or magnetic
proximity effect \cite{Eremeev.prb2013, Katmis.nat2016}, we use a
\emph{magnetic extension} of the TI surface (or a thin film) to
achieve an efficient time-reversal symmetry breaking.  Namely,
  taking advantage of the situation when the ferromagnetic and
  topological insulators have (i) exactly the same or an affine
  crystal structure and (ii) a similar atomic composition, we consider
  an FM insulator (FMI) film deposited on top of a TI surface. In this way, the
absence of a sharp interface between the two subsystems, minimal
differences in the atomic composition, and a perfect or a very good
crystal lattice matching make the FMI film a natural extension of the
  TI. As a consequence, the topological surface state substantially 
relocates into the FMI film and acquires a giant
DP gap (several tens of meV) due to a strong exchange interaction with
3$d$ moments, occupying a complete atomic layer in the
magnetically-extended region. This observation is supported by the
highly accurate total-energy calculations revealing an FM order with
an out-of-plane easy magnetization axis for a number of FMI/TI systems
under consideration. Moreover, the approach proposed here offers other
numerous advantages such as intrinsic Fermi level location inside the
gap, highly-ordered crystal structure, absence of the harmful trivial
surface or interface states, intactness of the TI bulk-like region,
and, finally, a lot of possible combinations of magnetic and
non-magnetic insulators. Altogether, these properties undoubtedly show
that the magnetic extension is an extremely promising way towards
enhancement of the QAHE observation temperature and eventual
realization of the quantized TME at TI surfaces without an external
magnetic field.

\section{Results and discussion} 

\subsection{Structure and magnetism} 

In the present work, we study systems consisting of 
tetradymite-type nonmagnetic semiconductor films of different
thicknesses (including those corresponding to 3D and 2D TIs),
sandwiched between two septuple-layer(SL)-thick films of a
tetradymite-family magnetic insulator. Such a system design is
experimentally feasible since the septuple or quintuple layers (QLs)
of the tetradymite-type compounds are very stable units on their own
due to the strong covalent-ionic type bonding between the atomic
layers inside the block. The stability of such blocks is confirmed by
a QL-by-QL epitaxial growth mode \cite{Zhang.nphys2010}. We
note that the geometry described above allows realizing both QAHE and
TME \cite{Wang.prb2015, Okada.ncomms2016}. Besides, 
the use of a thick TI slab allows visualizing the magnetic
extension effect in a full measure, when the topological surface state
 splits due to magnetism. On the other hand, a small
thickness of an FMI film ensures a near-surface location of the gapped
topological state even if it does not significantly penetrate into the extended 
region. Therefore, our theoretical predictions can be verified
experimentally by using a standard angle-resolved photoemission
spectroscopy. 

The starting point of our study is the fact
that the family of the tetradymite-like compounds
$X$B$^\mathrm{VI} \cdot $A$_2^\mathrm{V}$B$_3^\mathrm{VI}$
(A$^\mathrm{V}$ = Sb, Bi; B$^\mathrm{VI}$ = Se, Te; $X$ is, e.g., Ge, Sn or Pb)
\cite{Eremeev.ncomms2012, Silkin.jetpl2012} does not restrict itself
to the nonmagnetic materials only. These compounds crystallize in a
rhombohedral structure [space group R$\bar 3$m (166)], which
 is comprised of the SL building blocks stacked along the
$c$ axis and separated by a van der Waals gap. 
Importantly
for our purposes, there exist quite a few stable ${X}$B$^\mathrm{VI}$
compounds with the hexagonal low-index surfaces, what makes them
potentially compatible with those of the
A$_2^\mathrm{V}$B$_3^\mathrm{VI}$ tetradymite family. Among them, in
particular, there are magnetic semiconductors, an interesting
representative being the room temperature antiferromagnet MnTe
\cite{Podgorny.jpc1983}.  It is therefore quite logical to suggest the
MnB$^\mathrm{VI}\cdot $A$_2^\mathrm{V}$B$_3^\mathrm{VI}$ compounds
with the tetradymite-like structure to be stable as well. Recently,
this indeed has been confirmed by Lee et al. \cite{Lee.cec2013} who
reported a successful growth of the R$\bar 3$m-group bulk
MnBi$_2$Te$_4$ for the first time. The material was found to be a
$p$-type semiconductor and its in-plane lattice constant was measured
to be 4.334 \AA, which matches very well to that of various
tetradymite-type TIs like Bi$_2$Te$_3$, Sb$_2$Te$_3$ and others (see
Supplementary Note 1).  Having the same or allied crystal
structure, a very similar atomic composition and being
well-lattice-matched, such MnBi$_2$Te$_4$/TI (MBT/TI) systems are
expected to be readily grown by molecular-beam epitaxy. A natural
question then arises: how will the topological surface state of a
particular TI be changed upon a magnetic extension by an SL of
MnBi$_2$Te$_4$?

\begin{figure*}[!bth]
\begin{center}
\includegraphics[width=0.99\textwidth]{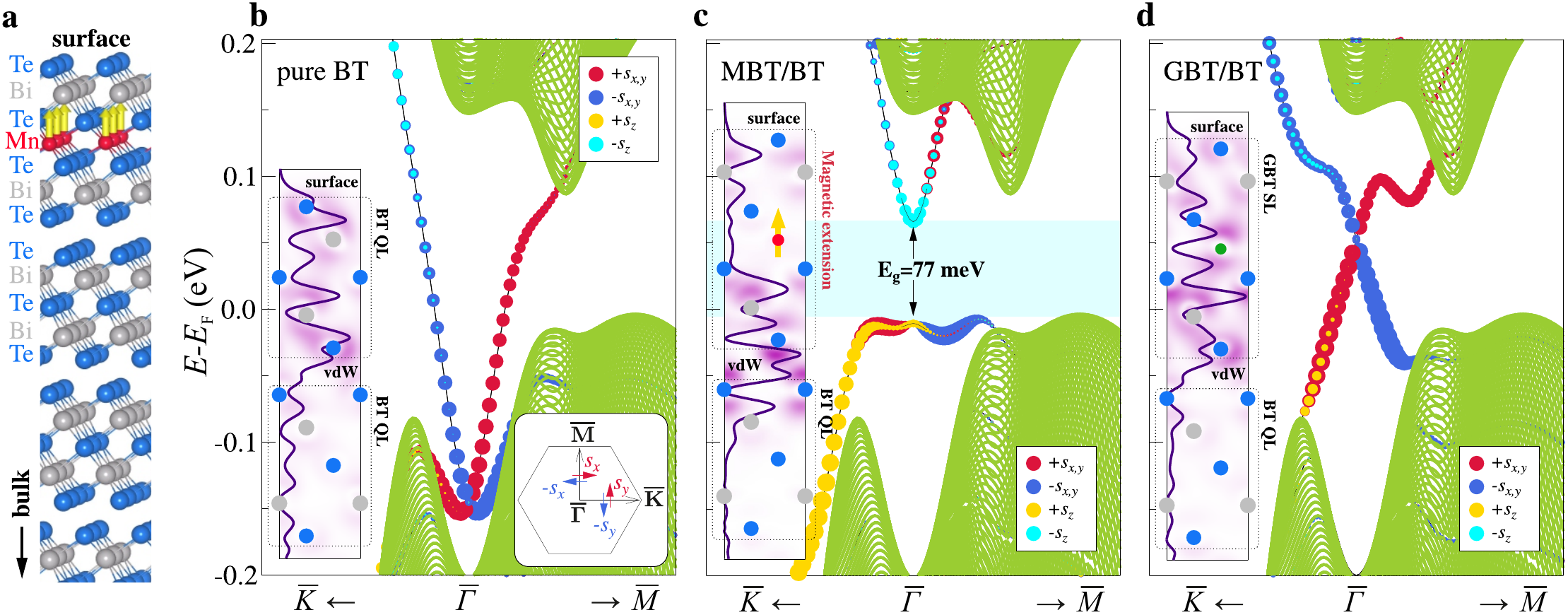}
\end{center}
\caption{(a) Atomic structure of MBT/BT with gold arrows indicating the Mn 3$d$ moments, 
pointing out of the surface plane. (b,c,d) Spin-resolved surface band spectra of BT (b), MBT/BT 
(c) and GeBi$_2$Te$_4$/BT (GBT/BT (d)). The size of color circles reflects the value and sign 
of the spin vector Cartesian projections, with red/blue colors corresponding to the positive/negative 
in-plane components (that directed perpendicular to the {\bf k} vector), and gold/cyan reflecting 
the out-of-plane components $+s_z$/$-s_z$. Green areas correspond to the bulk band structure 
projected onto the surface Brillouin zone. The insets to (b), (c) and (d) show the real space 
distribution of the topological surface state charge density as projected onto the $ac$-plane 
(color map in the white-to-pink palette) and as integrated over the $ab$ plane (purple line) for 
BT, MBT/BT and GBT/BT, respectively. In the insets, the 'vdW' stays for the van der Waals gap.} 
\label{fig:bt}
\end{figure*} 

Before answering this question, we first determine the magnetic ground
state of the MBT/TI systems. As a substrate for an MBT film
deposition, we take a thick film (6QL slab) of one of the most studied
TIs -- Bi$_2$Te$_3$ (BT), see Fig.~\ref{fig:bt}a. Our total energy
density functional theory calculations reveal that MBT/BT is a ferromagnet with an out-of-plane easy axis, Curie
temperature of 39 K, and a local moment of
4.613 $\mu_B$ per Mn atom (Supplementary Note 2). 

\subsection{Magnetically-extended TI surface for quantized magnetoelectric effect} 

We now study the surface band structure of the MBT/BT system.
Figure~\ref{fig:bt}b shows the topological state of Bi$_2$Te$_3$(0001)
with its DP located below the bulk valence band maximum
\cite{Sessi.prb2013}. The deposition of the MBT SL film on the BT
surface leads to substantial changes in the
low-energy spectrum. As it is seen in Fig.~\ref{fig:bt}c, the
surface band structure of MBT/BT is fully gapped -- the gap in
the DP reaches a gigantic value of 77 meV and the
Fermi level is located within it.  Moreover,
the surface band gap almost completely lies within the bulk one,
despite the fact that the original DP of the pure BT surface
is located
$\sim$150 meV lower than the bulk valence band maximum. A clear
explanation comes out if instead of the MBT/BT system the
[GeBi$_2$Te$_4$]$_\mathrm{1SL}$/Bi$_2$Te$_3$ (GBT/BT) one is
considered. While the GBT-extension of the Bi$_2$Te$_3$ surface
expectedly keeps the topological state gapless, it leads to a strong
upward shift of the DP which turns out to be located in the
fundamental band gap (Fig.~\ref{fig:bt}d). A similar behaviour has
recently been observed in the case of nonmagnetic TI-based heterostructures 
\cite{Menshchikova.nl2013, Men'shov.jpcm2014} and has been shown to depend on the electron 
affinities and band gap widths of the substrate and overlayer 
materials. The mechanism behind the giant DP
splitting is illustrated in the insets of Figs.~\ref{fig:bt}b,c,d,
showing the real space distribution of the topological surface states
of the respective cases. It can be seen, that the topological state
largely relocates into the magnetically extended region where the
time-reversal symmetry is broken owing to an out-of-plane
magnetization of the Mn layer. Despite the wave function of 
the topological surface state does not show a maximum on Mn atoms, 
its localization in the MBT SL and on the Mn layer in particular turns out 
to be sufficient to induce a very large splitting at the DP. In the Supplementary Note 3, we
show that the DP gaps up to 87 meV have been calculated for the MBT/BT
system within the physically-meaningful range of the effective
$U_{eff}$ parameter values, taking into account a strongly-correlated
nature of the Mn 3$d$-states. The ultimate value of this parameter
and, therefore, the DP gap size can be determined from photoemission
experiments that we hope to inspire.

The situation shown in Fig.~\ref{fig:bt}c is in stark contrast with
the case of well-defined interfaces, formed between magnetic and
topological insulators with different crystal structures, e.g. MnSe
and Bi$_2$Se$_3$. As shown in \cite{Eremeev.prb2013}{}, the Dirac cone
of Bi$_2$Se$_3$, localized in the topmost QL in the free surface case,
relocates to the underlying QL upon interfacing with MnSe, what
results in a quite moderate DP splitting of 8.5 meV. 
Furthermore, because of substantial modification of the interface potential, 
a trivial interface state appears, making the spectrum gapless.  In MBT/BT, 
the interface region is well comparable to the van der Waals region of the 
BT substrate and, therefore, the MBT/BT low-energy spectrum is
essentially free of parasitic trivial bands at any {\bf k}. Besides, the
atomic composition of the layers, lying near the van der Waals gap, 
for the MBT and interfacial BT QL essentially coincides up
to the third atomic layer inclusively, see Fig.~\ref{fig:bt}a. These
factors altogether lead to a magnetic extension of the Bi$_2$Te$_3$
surface characterized by the absence of the trivial states and by the
topological state that is largely localized 
in the magnetic block.

\begin{figure*}[!bth]
\begin{center}
\includegraphics[width=0.5\textwidth]{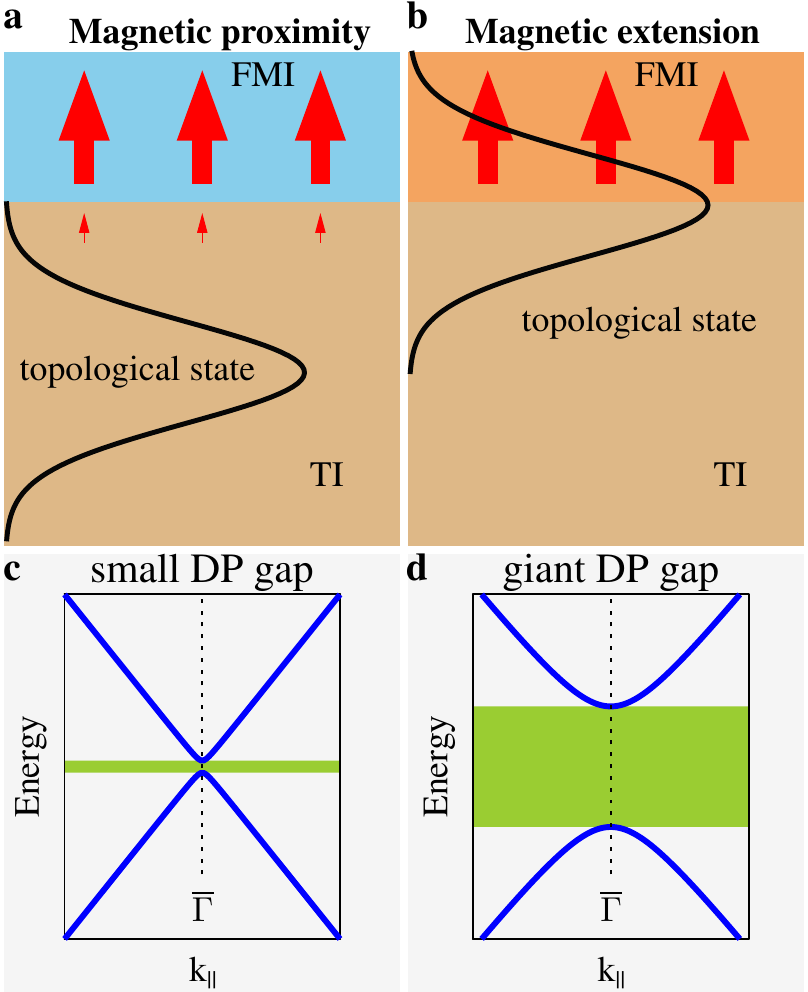}
\end{center}
\caption{
Schematic illustration of (a) the magnetic proximity effect following Ref.~\cite{Eremeev.prb2013}  
and (b) the magnetic extension of the topological insulator surface. The effects respectively yield small 
(c) and giant (d) DP gaps. Large and small arrows in panels (a) and (b) denote local and induced
magnetic moments, respectively.}
\label{fig:sketch}
\end{figure*}

It is worth to highlight a fundamental difference between the magnetic extension approach that 
we propose and the magnetic proximity effect, that has previously been used to lift the time-reversal
symmetry at the TI surface \cite{Eremeev.prb2013,Katmis.nat2016} or create a QAH state in graphene 
\cite{Qiao.prl2014, Zhang.srep2015}. In the magnetic proximity effect, a magnetically ordered 
system induces finite magnetization in a nonmagnetic system through the interface coupling {(Fig.~\ref{fig:sketch}a)}. 
The induced magnetization can lift band degeneracies in the nonmagnetic system: 
a small gap opens at the DP of a TI \cite{Eremeev.prb2013}, see Fig.~\ref{fig:sketch}c,  
or the Dirac cone of graphene experiences exchange splitting \cite{Qiao.prl2014, Zhang.srep2015}. 
In the magnetic extension approach the situation is completely different:
the topological state significantly penetrates into the magnetic film region 
and strongly splits due to direct interaction with the magnetic moments of Mn atoms (Fig.~\ref{fig:sketch}b,d).

{\renewcommand{\arraystretch}{1.5}%
\begin{table}[!bth]
\caption{
{Calculated values of the Dirac point band gaps for various systems}}
\label{tab:gaps} 
\begin{center}
\begin{tabular}%
{>{\PBS\centering\hspace{0pt}} p{4.50cm}%
>{\PBS\centering\hspace{0pt}} p{2.00cm}}
\hline
\hline
System & $E_g$ (meV) \\

\hline
\hline
MnBi$_2$Te$_4$/Bi$_2$Te$_3$ & 77 \\
MnBi$_2$Te$_4$/Sb$_2$Te$_3$ & 73 \\
MnBi$_2$Te$_4$/Bi$_2$Te$_2$Se & 52 \\
MnBi$_2$Te$_4$/Bi$_2$Te$_2$S & 47 \\
MnSb$_2$Te$_4$/Sb$_2$Te$_3$ & 25 \\
\hline
\hline
\end{tabular}
\end{center}
\end{table}
}

It is obvious from the above said that, due to its remarkable action on the topological state, 
magnetic extension appears to be a perfect approach for the quantized TME observation at 
the surfaces of thick TI films without magnetic field \cite{Wang.prb2015}. 
In Table~\ref{tab:gaps}, one can find some more examples of the systems with the 
giant DP splittings up to 73 meV. The corresponding band structures are shown in the 
Supplementary Note 4.
We envisage a hardly limited number of systems, which can be
produced using the magnetic extension approach. This is possible owing
to a great variety of the tetradymite-like TIs, consisting of
different blocks and their combinations \cite{Eremeev.ncomms2012, Silkin.jetpl2012},
disordered TIs of the Bi$_{2-x}$Sb$_{x}$Te$_{3-y}$Se$_y$(S$_y$) type,
TIs beyond the tetradymite family \cite{Kuroda.prl2010}, FMI/TI
superstructures and, possibly, other FMIs. 

\subsection{Thin films limit and QAH phase} 

\begin{figure*}[!bth]
\begin{center}
\includegraphics[width=0.99\textwidth]{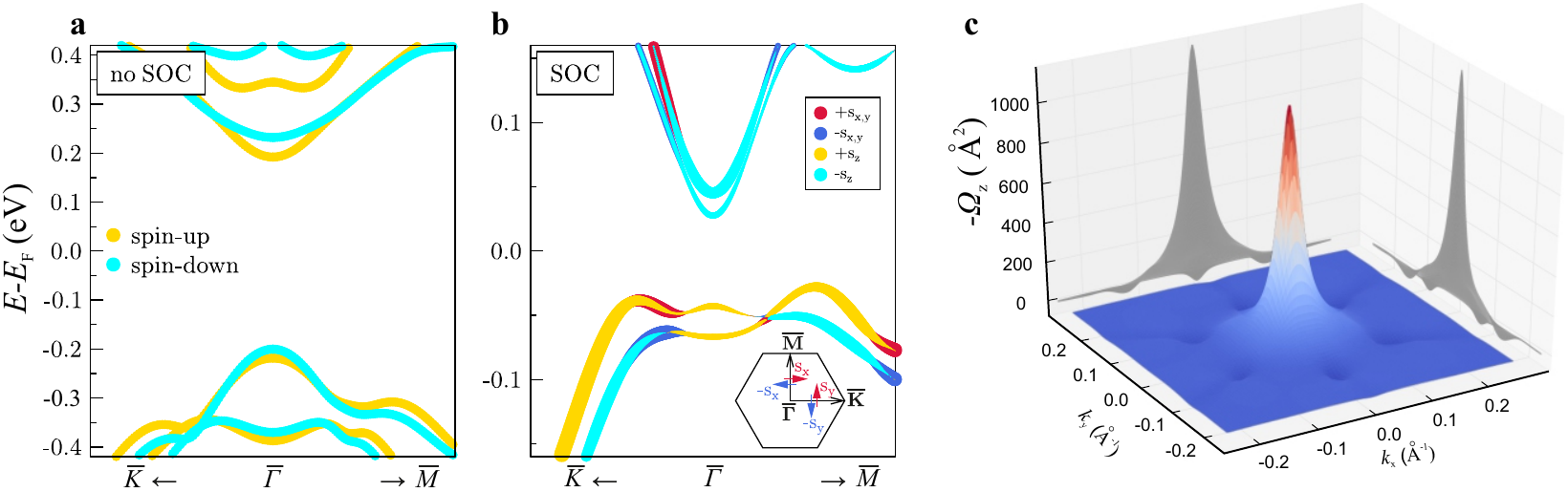}
\end{center}
\caption{
(a) and (b) Spin-resolved band structures of MBT/[Bi$_2$Te$_3$]$_{\mathrm{2QL}}$/MBT, 
calculated without and with SOC, respectively. The color coding is the same as in Fig.~\ref{fig:bt}. 
(c) Berry curvature calculated for MBT/[Bi$_2$Te$_3$]$_{\mathrm{2QL}}$/MBT.} 
\label{fig:qahe}
\end{figure*} 

We have constructed various MBT/[Bi$_2$Te$_3$]$_{n\mathrm{QL}}$/MBT
heterostructures with different number of the BT QLs, $n=1 - 5$.
Figure~\ref{fig:qahe} compares the band structures of the
MBT/[Bi$_2$Te$_3$]$_{\mathrm{2QL}}$/MBT sandwich calculated without
and with spin-orbit coupling included. Note that the calculations
have been done assuming the parallel alignment of magnetic moments of the
 MBT blocks separated by $n$ QL(s) of Bi$_2$Te$_3$. 
Since the MBT SLs are almost magnetically decoupled, 
this can always be achieved using an external magnetic field, as it has been	
done in the recent experiments showing the highest QAH temperatures in 
modulation-doped TI films \cite{Mogi.apl2015, Okada.ncomms2016} 
(see also Ref.~\cite{Men'shov.jetpl2016} for a theoretical analysis of 
the modulation-doped QAH insulator).
It can be seen in Figs.~\ref{fig:qahe}a,b that SOC
causes an inversion of a pair of spin polarized bands in the
$\overline{\Gamma}$-point vicinity, which is a prerequisite for the
QAH state onset. Indeed, when spin-orbit coupling is absent
(Fig.~\ref{fig:qahe}a), both the upper and lower gap edges are formed
by a pair of bands of opposite spins near the Brillouin zone
center. However, with SOC included, both upper valence bands (lower
conduction bands) feature the majority (minority) spins $+s_z$
($-s_z$) around the $\overline{\Gamma}$-point. In other words, one of
the two upper valence bands (lower conduction bands) changes its spin
on the opposite near $k_{||}=0$ owing to spin-orbit coupling, what can
be recognized by a change of the line's color in
Figs.~\ref{fig:qahe}a,b. To confirm the nontriviality of the
MBT/[Bi$_2$Te$_3$]$_{\mathrm{2QL}}$/MBT band structure in a more solid
way, the Chern number calculations have been performed. Using an
\emph{ab-initio}-based tight-binding approach (see Methods section), a
non-zero Berry curvature has been found near the
$\overline{\Gamma}$-point (Fig.~\ref{fig:qahe}c) yielding a Chern
number of $-1$ upon integration over the first Brillouin zone. This
result has been reproduced by an independent calculation of the Chern
number with the Z2Pack thus confirming the QAH state in
MBT/[Bi$_2$Te$_3$]$_{\mathrm{2QL}}$/MBT with a band gap of
56 meV. The equality of the Chern number $C$ to $-1$ for 
MBT/[Bi$_2$Te$_3$]$_{\mathrm{2QL}}$/MBT means that there will be a 
single dissipationless chiral edge mode at any border of the film. 
The Hall conductance facilitated by the edge mode is related to the 
Chern number as $\sigma_{xy}=C e^2/h$ and is therefore quantized to 
a value of $-e^2/h$ for the proposed system. 
Following the above-described strategy, we have found a number of QAH
insulators with the Chern number equal to $-1$ and the band gaps ranging 
from 16 to 61 meV (Table~\ref{tab:qahe_gaps} and Supplementary Note 5). 

{\renewcommand{\arraystretch}{1.5}%
\begin{table}[!bth]
\caption{{Calculated band gaps for various QAH insulators}}
\label{tab:qahe_gaps} 
\begin{center}
\begin{tabular}%
{>{\PBS\centering\hspace{0pt}} p{4.9cm}%
>{\PBS\centering\hspace{0pt}} p{1.9cm}}
\hline
\hline
System & $E_g$ (meV) \\

\hline
\hline
MBT/[Bi$_2$Te$_3$]$_{\mathrm{1QL}}$/MBT & 38 \\
MBT/[Bi$_2$Te$_3$]$_{\mathrm{2QL}}$/MBT & 56 \\
MBT/[Bi$_2$Te$_3$]$_{\mathrm{3QL}}$/MBT & 56 \\
MBT/[Bi$_2$Te$_3$]$_{\mathrm{4QL}}$/MBT & 60 \\
MBT/[Bi$_2$Te$_3$]$_{\mathrm{5QL}}$/MBT & 61 \\
MBT/[SnSb$_2$Te$_4$]$_{\mathrm{1SL}}$/MBT & 40 \\
MBT/[SnSb$_2$Te$_4$]$_{\mathrm{2SL}}$/MBT & 16 \\
MBT/[PbSb$_2$Te$_4$]$_{\mathrm{1SL}}$/MBT & 55 \\
MBT/[PbSb$_2$Te$_4$]$_{\mathrm{2SL}}$/MBT & 55 \\
\hline
\hline
\end{tabular}
\end{center}
\end{table}
}

\section{Outlook and conclusions} 
The above-described properties reveal that the magnetic extension is
an extremely promising way of the time-reversal symmetry breaking in
TIs. In the Supplementary Note 6 we give an extensive comparison to
other approaches of the time-reversal symmetry breaking in TIs
\cite{Liu.prl2009, Lee.pnas2015, Henk.prl2012.2, Eremeev.prb2013} as
well as to specific QAH systems (both synthesized
\cite{Checkelsky.natp2014, Kou.prl2014, Chang.natm2015,
  Kou.ncomms2015, Feng.prl2015, Mogi.apl2015, Lachman.sciadv2015,
  Grauer.prb2015, Okada.ncomms2016, Chang.prl2016} and
theoretically-proposed \cite{Liu.prl2008, Xu.prl2011, Zhang.prl2012,
  Wang.prl2013, Zhang.srep2013, Garrity.prl2013, Garrity.prb2014,
  Wang.prl2014, Wu.prl2014, Zhang.prl2014, Jin.srep2015, Zhou.nl2015,
  Liu.prb2015, Xu.nl2015, Xu.prl2015, Dong.prl2016, Chen.srep2016,
  Menghao.2dmat}) that is summarized here as follows.  First, in
comparison to the strongly-disordered
(V$_x$)Cr$_{x}$(Bi$_{1-y}$Sb$_{y}$)$_{2-x}$Te$_3$ QAH insulators
studied in the experiments \cite{Checkelsky.natp2014, Kou.prl2014,
  Chang.natm2015, Kou.ncomms2015, Feng.prl2015, Mogi.apl2015,
  Lachman.sciadv2015, Grauer.prb2015, Okada.ncomms2016,
  Chang.prl2016}, the MBT/TI/MBT sandwiches yield up to at least six
times larger band gaps and are expected to show a highly-ordered structure,
inherent of stoichiometric materials. The latter makes the magnetic
extension based systems stable against the in-gap dopant states
\cite{Sanchez-Barriga.ncomms2016} as well as possible
superparamagnetic behavior \cite{Lachman.sciadv2015,
  Grauer.prb2015}. We speculate that an increase of nearly one order
of magnitude of the QAHE observation temperature may be expected owing
to these advances.  Second, having considered a variety of theoretical
proposals of the QAH insulators beyond those of the tetradymite family
\cite{Liu.prl2008, Xu.prl2011, Zhang.prl2012, Wang.prl2013,
  Zhang.srep2013, Garrity.prl2013, Garrity.prb2014, Wang.prl2014,
  Wu.prl2014, Zhang.prl2014, Jin.srep2015, Zhou.nl2015, Liu.prb2015,
  Xu.nl2015, Xu.prl2015, Dong.prl2016, Chen.srep2016}, one can
conclude that the QAH insulators based on the magnetic extension
approach proposed appear nowadays as an optimal platform for the QAHE.
Indeed, while the techniques for a controllable growth of the wide
band gap honeycomb-structure QAH insulators \cite{Wu.prl2014,
  Garrity.prl2013} are still to be elaborated, the
magnetic-extension-based QAH insulators can be implemented immediately
making use of existing advanced technology of the tetradymite-type
compounds epitaxial growth. 
Finally, the magnetic extension approach is expected to partly circumvent
the difficulty coming from the constraint on the photon frequency
$\omega \ll E_g/\hbar$ \cite{Qi.prb2008} ($E_g$ is the DP gap), that requires
wide gap materials for an accurate measurement of the topological 
Kerr and Faraday rotations (see Supplementary Note 6). We stress that
among the experimentally-feasible specific proposals, magnetically-extended 
TI surfaces feature largest DP gaps, what makes them best candidates for 
realization of the quantized TME at TI surfaces at zero magnetic field.

Thus, using \emph{ab initio} band structure calculations, we have
proposed a magnetic extension of topological insulator surfaces -- a
novel approach for the time-reversal symmetry breaking. The key idea
behind it, is the use of topological and magnetic insulators of the
allied crystal structure and similar atomic composition, such that the
surface features of the former are naturally extended upon the
deposition of the latter. In this case, the topological surface state
does not meet any significant interfacial potential and largely
penetrates into the magnetically-extended part getting gigantically
split, once the ferromagnetic state with an out-of-plane magnetization
onsets there. Moreover, in such systems, the Fermi level is
intrinsically located within the induced Dirac point gap, while 
trivial surface or interface states are essentially absent in the
low-energy spectrum. Importantly, the approach relies on the use of
the stoichiometric magnetic compounds thus ruling out possible
disorder-related effects.  Such a combination of properties renders
the magnetically-extended topological insulator surface to be a unique
system, perfectly suitable for the eventual experimental observation
of the topological magnetoelectric effect at TI surfaces without
magnetic field, as well as for the realization of the quantum
anomalous Hall state significantly beyond temperatures reached to
date.

\section{Methods} 

Electronic structure calculations were carried out within the density
functional theory using the projector augmented-wave method
\cite{Blochl.prb1994} as implemented in the VASP code \cite{vasp1,
  vasp2}. The exchange-correlation energy was treated using the
generalized gradient approximation \cite{Perdew.prl1996}. The
Hamiltonian contained the scalar relativistic corrections and the
spin-orbit coupling was taken into account by the second variation
method \cite{Koelling.jpc1977}. In order to describe the van der Waals
interactions we made use of the DFT-D2 \cite{Grimme.jcc2006} and the
DFT-D3 \cite{Grimme.jcp2010, Grimme.jcc2011} approaches, giving
similar results. The energy cutoff for the plane-wave expansion was
set to 270 eV. The Mn $3d$-states were treated employing the GGA$+U$
approach \cite{Anisimov1991} within the Dudarev scheme
\cite{Dudarev.prb1998}. Taking into account that the Mn layer local
environment in MBT is the same as that in bulk MnTe, the $U_{eff}=U-J$
value for the Mn 3$d$-states was chosen to be the same as the one
estimated for the bulk MnTe case \cite{Youn2005},
i.e. 5.34~eV. Nevertheless, an extensive testing was performed for the
MBT/BT system in order to ensure stability of the results against the
$U_{eff}$ value (see Supplementary Note 3).  It was found that the
magnetic ordering and magnetic anisotropy do not change qualitatively
when the $U_{eff}$ value changes from 3 to 5.34 eV.

The magnetically-extended TI surfaces were simulated within a model of
repeating slabs separated by a vacuum gap of a minimum of 10~\AA. The
FMI films were symmetrically attached to both sides of the substrate
slab to preserve the inversion symmetry, which was maintained for the
magnetically-ordered cases as well. Our total-energy calculations
show, that the lateral location of the FMI SL that  maintains the 
$...ABCABCABC...$ stacking of the substrate layers is the most favorable 
one. Therefore all calculations have been performed for this type of
connection between the substrate and FMI film.
The thicknesses of the TI
substrates chosen were such that the maximal hybridization gap in the
DP did not exceed 1 meV. These were the 6 QL and 7 SL slabs for the
A$_2^\mathrm{V}$B$_3^\mathrm{VI}$ and
A$^{\mathrm{IV}}$B$^\mathrm{VI}\cdot
$A$_2^\mathrm{V}$B$_3^\mathrm{VI}$
TIs, respectively. The in-plane lattice parameters of the
magnetically-extended TIs were fixed to the experimental ones of
corresponding TIs, while the interlayer distances were optimized for
the two upper structural blocks using a conjugate-gradient algorithm
and a force tolerance criterion for convergence of 0.03 eV/{\AA}
{}(spin-orbit coupling was included during the
relaxation). Relaxations and electronic structure calculations were
performed using a $\overline \Gamma$-centered $k$-point grid of
$11\times 11\times 1$ in the two-dimensional Brillouin zone.

To model the FM and collinear AFM phases, the ($1 \times \sqrt{3}$)
rectangular in-plane supercells containing two atoms per atomic layer
were constructed for each system under consideration. For the
verification of the MBT/BT FM state stability against the
non-collinear AFM state formation the
$(\sqrt{3}\times\sqrt{3})R30^\circ$ in-plane supercells containing
three atoms per atomic layer were used. Both ($1 \times \sqrt{3}$) and
$(\sqrt{3}\times\sqrt{3})R30^\circ$ in-plane cells as well as the
magnetic structures mentioned are visualized in the Supplementary Note
2. In all these calculations, the slab thicknesses were maintained the
same as for the surface band structure calculations. For the
total-energy calculations, the two-dimensional Brillouin zones were
sampled by the $9\times 5\times 1$ and $5\times 5\times 1$
$\overline \Gamma$-centered $k$-point grids in the cases of
($1 \times \sqrt{3}$) and $(\sqrt{3}\times\sqrt{3})R30^\circ$ in-plane
cells, respectively.

For the calculation of the band contribution to the magnetic
anisotropy energy, $E_b$, the $k$-mesh of $25\times 25\times 1$ was
chosen after the convergence tests performed (see Supplementary Note
7).  The calculations were done for the thicknesses of 44 and 56
atomic layers for the magnetically-extended surfaces of the QL- and
SL-based TIs, respectively.  To calculate $E_b$, the energies for
three inequivalent magnetization directions -- Cartesian $x$, $y$
(in-plane) and $z$ (out-of-plane) -- were calculated and $E_b$ was
determined as the difference $E_{\mathrm{in-plane}}$ -- $E_z$, where
the $E_{\mathrm{in-plane}}$ is the energy of the most energetically
favorable in-plane direction of magnetization. The energy convergence
criterion was set to 10$^{-7}$ eV providing a well-converged $E_b$
(the values of the order of few tenth of meV) while excluding
"accidental" convergence. The cutoff radius of a minimum of 20 microns
was used to calculate dipole-dipole contribution, $E_d$, to the
magnetic anisotropy energy, $E_a$.

Exchange coupling parameters of the MnBi$_2$Te$_4$ SL were calculated
using the Korringa-Kohn-Rostoker method within a full potential
approximation to the crystal potential \cite{Geilhufe2015}. We took
an angular momentum cutoff of $l_{max} = 3$ for the Green's function
and a k-point mesh of $25 \times 25 \times 1$ for the 2D Brillouin
zone integration.

The Chern numbers have been independently calculated using Z2Pack
\cite{Soluyanov.prb2011, Gresch.arxiv2016} and \emph{ab-initio}-based
tight-binding calculations within the WANNIER90 interface to the VASP
~\cite{marzari_vanderbilt1997,Mostofi2008}.  In the latter case, using
a tight-binding Hamiltonian $H$ in the Wannier function basis, the
Berry curvature tensor has been calculated applying Kubo
formula~\cite{Thouless.prl1982,Yao.prl2004}:
\begin{equation}
\Omega_{n,\alpha\beta}(\mathbf{k})= -\mathrm{Im} \sum_{v \ne n} \frac{< n | \nabla_{\alpha} H(\mathbf{k}) |v> <v| \nabla_{\beta} H(\mathbf{k}) | n >}{(E_n(\mathbf{k})-E_v(\mathbf{k}))^2},
\end{equation}
where $\nabla_{\alpha} H$ is a velocity operator, while $| n >$ and
$| v >$ are the Bloch states for each $\mathbf{k}$ with energies
$E_n(\mathbf{k})$ and $E_v(\mathbf{k})$, respectively.  Total Berry
curvature has been obtained by summation over the occupied states
$\Omega_{\alpha\beta}(\mathbf{k}) = \sum_{n} f_n
\Omega_{n,\alpha\beta}(\mathbf{k})$,
$f_n$ being the Fermi-Dirac distribution.  Three components of the
"magnetic" gauge field are obtained using
$\Omega_{\gamma}(\mathbf{k})=\epsilon_{\alpha\beta\gamma}
\Omega_{\alpha\beta}(\mathbf{k})$.
Here $\epsilon_{\alpha\beta\gamma}$ is a three-component antisymmetric
Levi-Civita tensor.  Chern number is evaluated as a Berry gauge flux
over the 2D Brillouin zone,
$C=1/(2\pi)\int_S \Omega_z(\mathbf{k}) d{k^2}$, where $S$ is the 2D
Brillouin zone area and $\Omega_z(\mathbf{k})$~-- a normal component
of the Berry curvature.

\section{Acknowledgments}

M.M.O. acknowledges useful discussions with S.V. Eremeev.  We
acknowledge support by the University of the Basque Country (Grant
Nos. GIC07IT36607 and IT-756-13), the Spanish Ministry of Science and
Innovation (Grant Nos. FIS2013-48286-C02-02-P, FIS2013-48286-C02-01-P,
and FIS2016-75862-P) and Tomsk State University Academic
D.I. Mendeleev Fund Program in 2015 (research grant N
8.1.05.2015). Partial support by the Saint Petersburg State University
project No.  15.61.202.2015 is also acknowledged. A.E. acknowledges
financial support from DFG through priority program SPP1666
(Topological Insulators). The calculations were performed in the
Donostia International Physics Center and 
Resource Center "Computer Center of SPbU" (http://cc.spbu.ru).

\section*{References}

\providecommand{\newblock}{}

\end{document}